\documentclass{an}
\usepackage{graphicx}
\usepackage{times}
\usepackage{fancyhdr}
\begin{document}

\title{CTK - A new CCD Camera at the University Observatory Jena\thanks{Based on observations obtained with telescopes of the University Observatory Jena, which
is operated by the Astrophysical Institute of the Friedrich-Schiller-University.}}

\author{Markus~Mugrauer} \institute{Astrophysikalisches
Institut und Universit\"{a}ts-Sternwarte Jena, Schillerg\"{a}{\ss}chen 2-3, 07745 Jena, Germany}

\date{Received; accepted; published online}

\abstract{The Cassegrain-Teleskop-Kamera (CTK) is a new CCD imager which is operated at the
University Observatory Jena since begin of 2006. This article describes the main characteristics of
the new camera. The properties of the CCD detector, the CTK image quality, as well as its detection
limits for all filters are presented. }
\correspondence{markus@astro.uni-jena.de}

\maketitle

\section{Introduction}

The University Observatory Jena is located close to the small village Gro{\ss}schwabhausen, west of the
city of Jena. The Friedrich Schiller University operates there a 0.9\,m reflector telescope which
is installed at a folk mount (see e.g. Pfau 1984)\nocite{pfau1984}. The telescope can be used
either as a Schmidt-Camera, or as a Nasmyth telescope. In the Schmidt-mode ($f/D=3$) the telescope
aperture is limited to the aperture of the installed Schmidt-plate $D=0.6$\,m. In the Nasmyth-mode
the full telescope aperture $D=0.9$\,m is used at $f/D=15$.

In March 2006 a new CCD imager saw first light at the University Observatory Jena. The new
\textbf{\large C}assegrain-\textbf{\large T}eleskop-\textbf{\large K}amera (CTK) is installed at
the $D=0.25$\,m auxiliary Cassegrain telescope ($f/D=9$), which is mounted at the tube of the
0.9\,m telescope. The CTK is operated from the control room which is located in the first floor of
the observatory under the telescope dome.

During the first months in 2006 the CTK characteristics were studied (dark current, bias stability,
linearity, detection limits), and first test observations (photo- and astrometric measurements)
were carried out. From end of 2006 on the CTK was used then for several scientific projects, mainly
photometric programs to study the variability of young stars, follow-up observations of
long-periodical variable stars, as well as of stars with transiting planets.

In this paper I present the main characteristics of the CTK. In the second section all components
of the camera are described. Section 3 summarizes all properties of the CTK detector. The CTK image
quality is discussed then in section 4. The last two sections describe the quantum efficiency of
the CTK detector, as well as the detection limits achieved with the new imaging camera.

\begin{figure}[h!]
\resizebox{\hsize}{!}{\includegraphics[angle=-90]{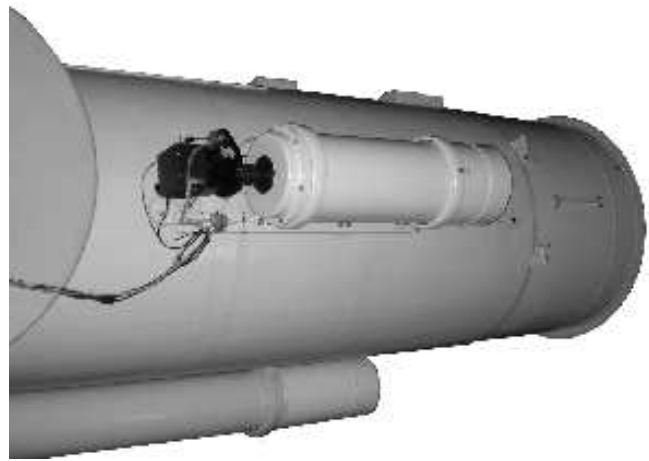}} \caption{This image shows the tube of
the 0.9\,m reflector telescope of the University Observatory Jena. The 0.25\,m Cassegrain auxiliary
telescope is mounted at the tube of the 0.9\,m reflector. The Cassegrain-Teleskop-Kamera (CTK) is
installed at the Cassegrain telescope. A detailed view on all individual components of the CTK is
shown in Fig.\,\ref{ctk2}.} \label{ctk1}
\end{figure}

\section{CTK components}

\begin{figure}[h]
\resizebox{\hsize}{!}{\includegraphics[]{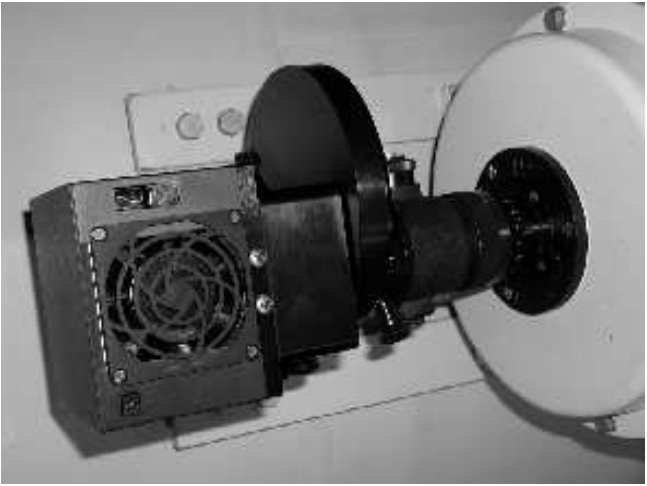}} \caption{The Cassegrain-Teleskop-Kamera (CTK)
at the 0.25\,m auxiliary Cassegrain telescope. This image shows a close view on all individual
components of the CTK. The camera is equipped with a manual focuser installed directly at the
backside of the Cassegrain telescope, followed by a five position filter wheel, and the CTK camera
head with the big cooling housing at its backside.} \label{ctk2}
\end{figure}

As it is shown in Fig.\,\ref{ctk2} the CTK is directly connected to the 0.25\,m Cassegrain
telescope with a manual focuser which also allows a rotation of the camera around the optical axis.
Between the focuser and the camera head a filter wheel is mounted which offers five filter slots
each with a diameter of 50\,mm. The filters B, V, R, and I from the Bessel standard system
(\cite{bessel1990}), as well as z$'$ from the SDSS system (\cite{fukugita1996}) are used. All
filters exhibit comparable thicknesses which guarantees only a small variation of the focal plane
in all filters. At the backside of the filter wheel the CTK camera head is located with its big
cooling housing. The CTK camera head contains the CCD detector, as well as the readout and cooling
electronics. The CTK detector is installed in an insulated cell with an entrance window in front of
the detector. The detector cell is dried and filled with Argon gas for suppression of humidity. An
iris shutter with four shutter blades is installed in front of the entrance window of the detector
cell. The shutter can be opened for integration times down to 0.05\,s. In the big cooling housing
at the backside of the camera head a fan removes warm air from the hot side of the Peltier cooling
device.

\section{CTK Detector}

The CCD sensor of the CTK is a back-illuminated SITe TK1024 of grade 2. The multi-pined-phased
technology (MPP) of this detector reduces its dark current also at higher cooling temperatures
above -50\,$^{\circ}$C. Furthermore, it minimizes surface residual image effects. The sensor size
measures 24.6\,mm$\times$24.6\,mm and is composed of 1024$\times$1024 pixel, each with a size of
24\,$\mu$m$\times$24\,$\mu$m.

\begin{figure}[h]
\resizebox{\hsize}{!}{\includegraphics[]{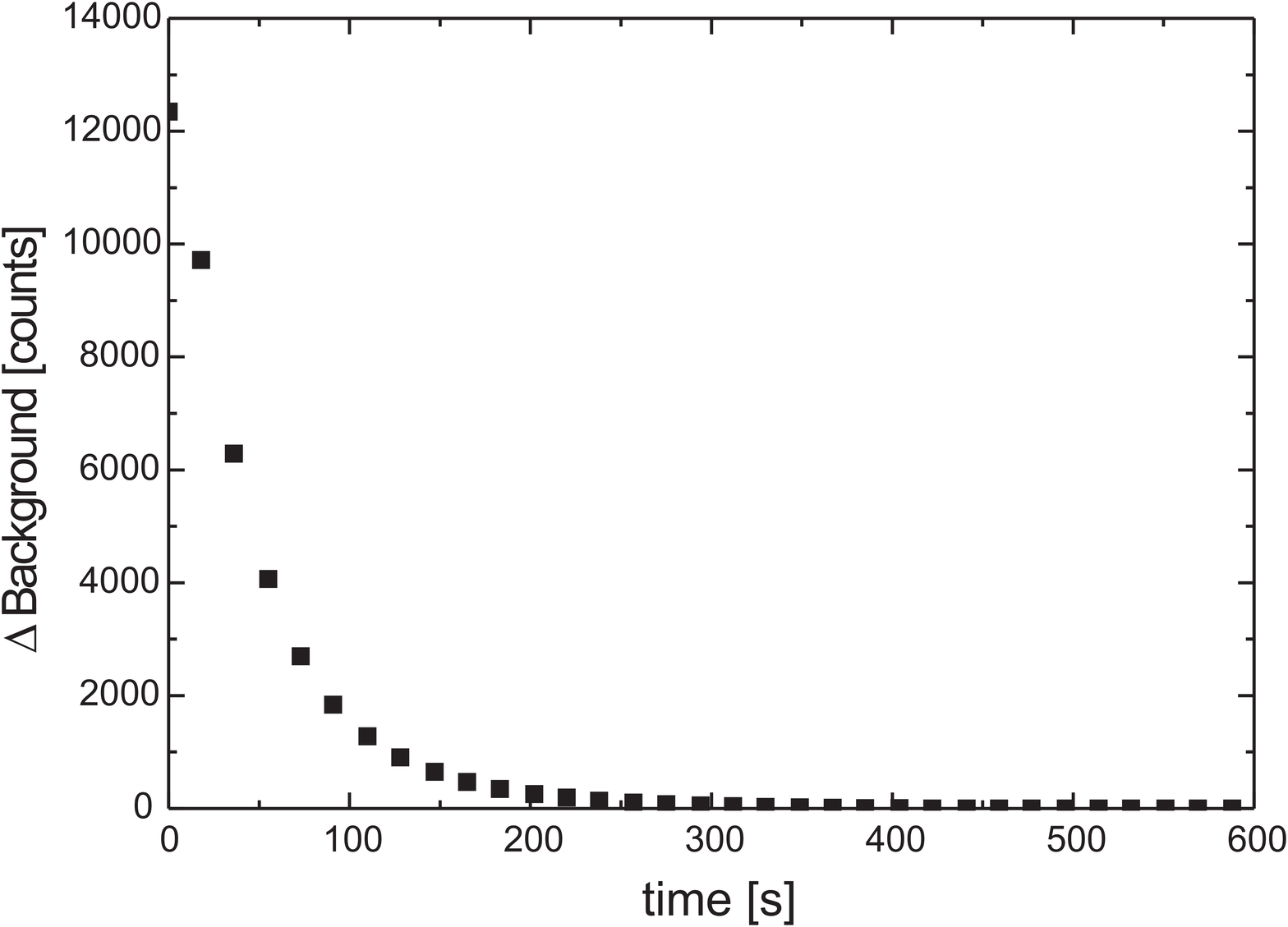}} \caption{Cool-down of the CTK detector. The
integrated cooling and temperature stabilization electronic of the CTK allows a sensor cooling down
to 42\,K below ambient air temperature. This plot shows the difference of the detector background
signal to its value after detector cooling and temperature stabilization is completed. Begin of the
cooling process is at $t=0$\,s. After already 7\,min the CTK background signal does not decrease
significantly anymore, and after 10\,min the background signal is stabilized.} \label{cooldown}
\end{figure}

The CTK sensor is Peltier cooled and a temperature difference of down to 42\,K below ambient air
temperature can be reached. The cooling of the CCD sensor is controlled by an integrated cooling
electronic. After the cooling temperature is selected the cooling electronic cools down the
detector and stabilizes its temperature at the given value. After the start of the detector cooling
about 10\,min are needed until the chosen cooling temperature is reached and stabilized. The
typical detector cool-down was measured and is shown in Fig.\,\ref{cooldown}. After the begin of
the sensor cooling the measured background signal of the CTK detector decreases exponentially.
After already 7\,min it does not change significantly, and after 10\,min the background signal is
stabilized.

The camera readout is done serial with a readout and data transfer rate of 43.7\,kpixel/s and a
typical read out noise of 7\,$e^{-}$. The analog-digital-converter (ADC) of the readout electronic
works at 16\,bit, i.e. the file size of a CTK image (1024$\times$1024 pixel) is about 2.1\,MB. Due
to the MPP technology the full well capacity of the CTK detector is reduced to 150000\,e$^{-}$.
With an averaged bias level of 14900\,counts and a bit depth of 16bit this yields a detector gain
of 3\,e$^{-}$/count.

\begin{table}[h!] \centering\caption{CTK dark current for a range of detector cooling temperatures.
The dark current remains on a low level for cooling temperatures below about -~25$^{\circ}$C, the
so-called critical cooling temperature. Operation of the CTK above this temperature limit should be
avoided due to a significantly increased dark current.} \label{table_darktemp}
\begin{tabular}{cc|cc}\hline
T & dark current & T & dark current\\
$[^{\circ}$C] & [counts/min] & $[^{\circ}$C] & [counts/min]\\
\hline
$-35$ & $14 \pm 4$   & $-10$    &  $690 \pm 38$\\
$-30$ & $31 \pm 5$   & $-05$    & $1459 \pm 65$\,\,\,\\
$-25$ & $68 \pm 8$   & $~~~00$  & $3095 \pm 124$\\
$-20$ & $149 \pm 13$ & $+05$    & $6407 \pm 195$\\
$-15$ & $323 \pm 23$ &          & \\
\hline\hline
\end{tabular}
\end{table}

\begin{figure}[h]
\resizebox{\hsize}{!}{\includegraphics[]{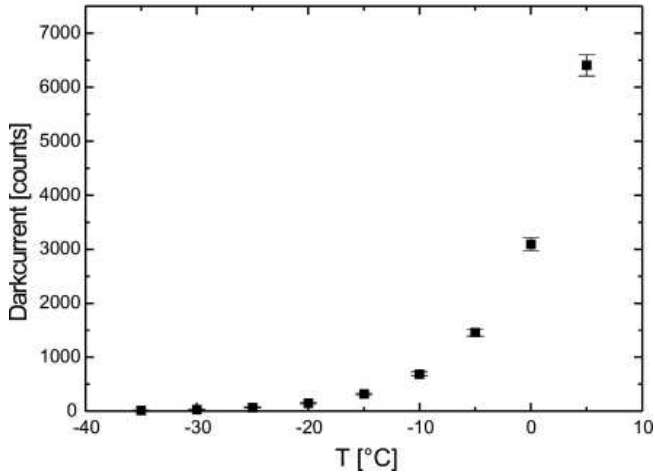}} \caption{The CTK dark current for an
integration time of 60\,s measured at different detector cooling temperatures. The dark current
remains on a low level below the critical cooling temperature of $-25\,^{\circ}$C.}
\label{darktemp}
\end{figure}

The measured dark current for a range of cooling temperatures is summarized in
Tab.\,\ref{table_darktemp}, and illustrated in Fig.\,\ref{darktemp}. The dark current scales
exponentially with the cooling temperature and remains low below a critical cooling temperature of
$-25\,^{\circ}$C while it significantly increases for higher temperatures. Therefore, the CTK
detector should always be operated below the critical cooling temperature of about
$-25\,^{\circ}$C. The integrated Peltier cooling guarantees operation below the critical
temperature limit during all nights of a year. In cold winter nights cooling temperatures down to
$-60\,^{\circ}$C can be reached. In typical warm summer nights cooling temperatures of about
-25\,$^{\circ}$C also yield low dark currents.

\begin{figure}[h]
\resizebox{\hsize}{!}{\includegraphics[]{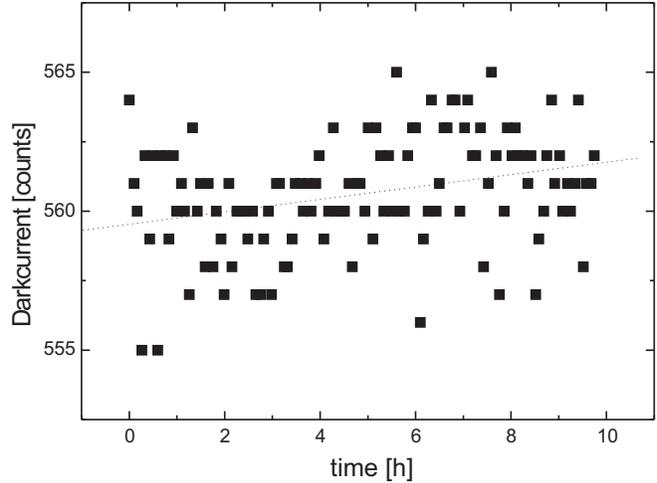}} \caption{The CTK dark current for 60\,s
of integration time monitored for a range of time. The chosen detector cooling temperature is
$-11\,^{\circ}$C. The dark current does not vary significantly over time. There is only a
negligible increase of 0.2\,counts/h observed which is illustrated in the plot with a dotted line.}
\label{darkcurrent}
\end{figure}

Due to the installed temperature stabilization electronic the detector temperature and hence also
the dark current is highly stable throughout the night. The typical dark current stability of the
CTK detector during a night was measured and is illustrated in Fig.\,\ref{darkcurrent}. The dark
current does not show any significant drift during the night.

\begin{figure}[h!]
\resizebox{\hsize}{!}{\includegraphics[]{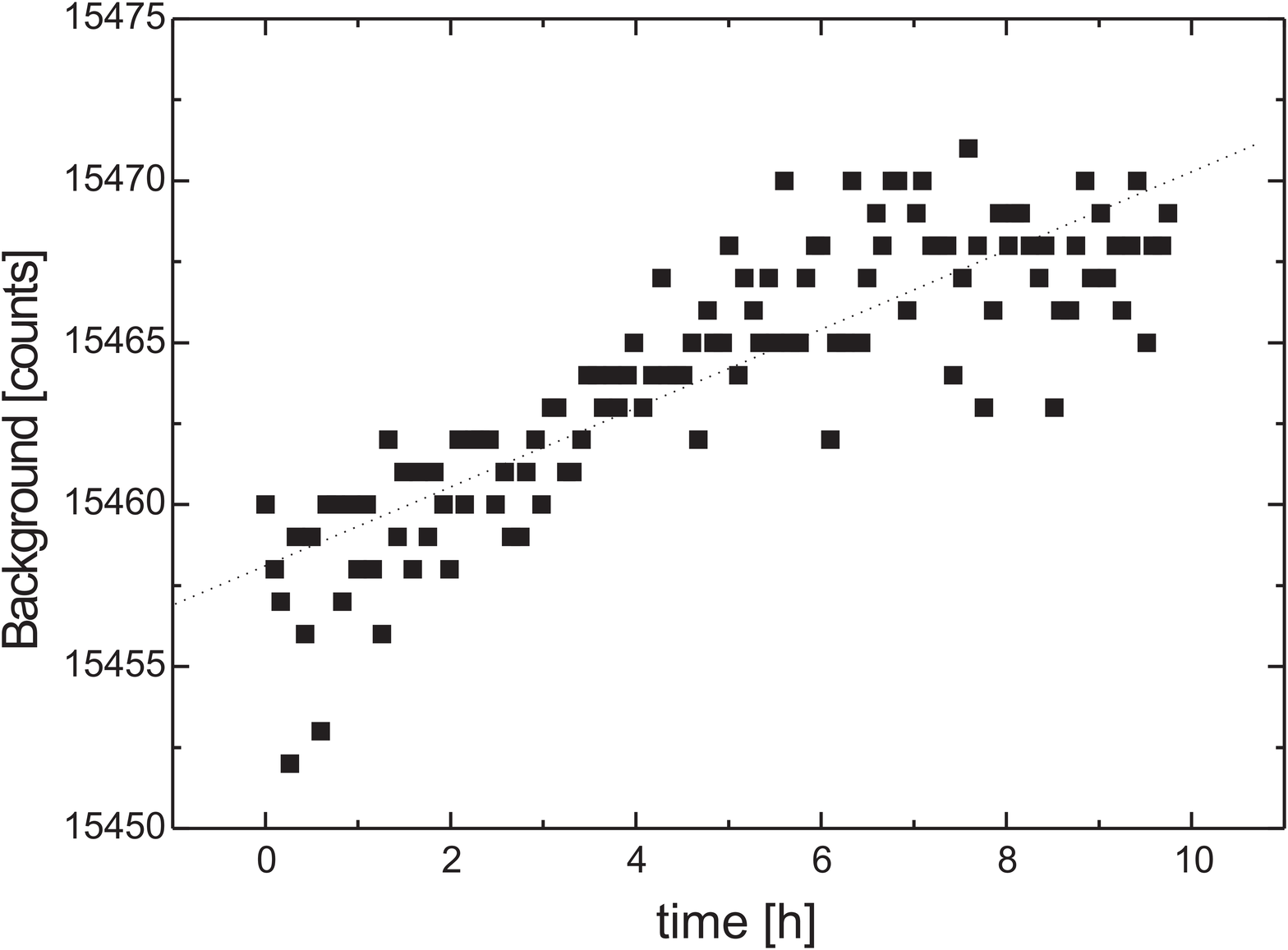}} \caption{The CTK detector background
signal measured for a range of time. During the monitoring the detector cooling temperature is
stabilized at -11\,$^{\circ}$C. The integration time is 1\,min. Although the dark current is stable
over time (see Fig.\,\ref{darkcurrent}) a change in the background signal up to 2\,counts/h is
observed (see dotted line). This change is caused by a drift of the detector bias level which is
dependant on the ambient temperature. The typical temperature drift of the bias is 5\,counts/K.}
\label{background}
\end{figure}

However, as it is shown in Fig.\,\ref{background} the CTK background signal (bias + dark current)
slightly varies with a typical change of up to 2\,counts/h. This drift is caused by a change of the
detector bias level which is dependant on the ambient temperature of the readout electronic. The
drift of the bias level due to a change of temperature is found to be 5\,counts/K with an averaged
bias level of 14900\,counts for typical CCD operation. In order to correct this drift it is
recommended to take some few bias calibration images throughout the night.

\begin{figure}[h!]
\resizebox{\hsize}{!}{\includegraphics[]{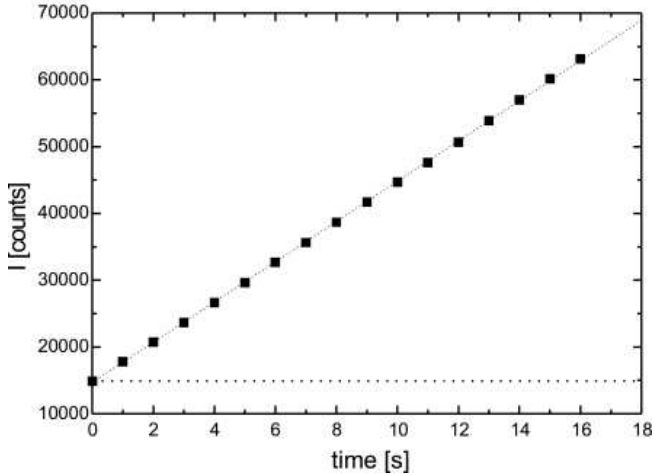}} \caption{The linearity of the CTK
detector. The detected signal $I$ of a domeflat is plotted for a range of integration time. A quite
linear dependency (closely dotted line) between integration time and background signal is found for
the whole dynamic range from the bias level (widely dotted line) at 14900\,counts in average, up to
the upper limit of the CTK ADC.} \label{linearity}
\end{figure}

\begin{figure}[h]
\resizebox{\hsize}{!}{\includegraphics[]{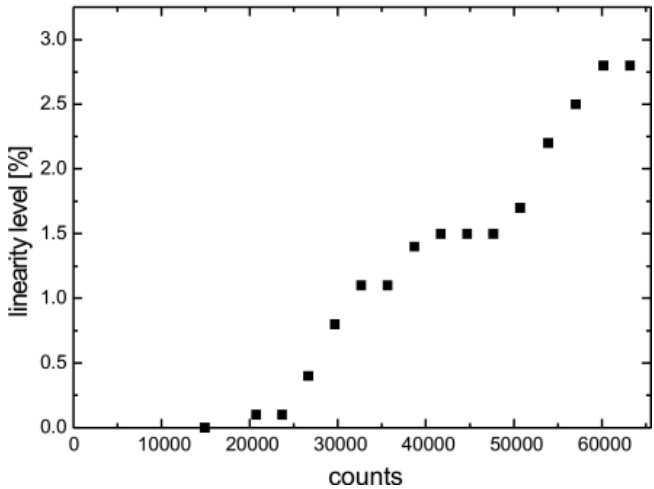}} \caption{The lineartity level of the
CTK detector for a range of background signal. The detector is linear on a 3.5\,\% level over the
full dynamic range from the bias level at 14900\,counts in average up the upper limit of the CTK
ADC.} \label{linearitylevel}
\end{figure}

In order to measure the linearity of the CTK sensor domeflats with different integrations times at
a constant flat illumination were taken. The found dependency of the measured detector intensity
$I$ in unit of counts for a range of integration time is illustrated in Fig.\,\ref{linearity}. The
detector is quite linear over its full dynamic range from the bias level at 14900\,counts in
average up to the upper limit of the ADC. The linearity level of the CTK detector, i.e. the
deviation of the slope compared to the slope measured for an intensity between about 15000 and
18000\,counts is shown in Fig.\,\ref{linearitylevel}. The CTK detector exhibits a high level of
linearity (better than 1\,\%) below an intensity of 30000\,counts. The linearity level increase
slightly for higher intensities and reaches 2.5\,\% at an intensity of 63000\,counts, which also
does not change significantly up to the upper limit of the CTK ADC. Hence, precise photometric
measurements are feasible with the CTK spanning its complete dynamic range.

\section{CTK Astrometry and Image Quality}

The 24\,$\mu$m pixel of the CTK detector exhibit a pixel scale of 2.2065$\pm$0.0008\,$''$/pixel at
the 0.25\,m Cassegrain ($f/D=9$), yielding 37.7$'$$\times$37.7$'$ field of view. The pixel scale is
determined with several CTK images, which are astrometrically calibrated using reference sources
listed in the 2MASS point source catalogue (\cite{skrutskie2006}). The CTK manual focuser allows a
full rotation of the camera head around the optical axis. Therewith, one can obtain a precise north
alignment of the CTK detector. As it was shown in a series of tests a north alignment precision
better than 0.1$^{\circ}$ can be achieved.

Due to a slightly different thickness of the individual CTK filters the focal plane slightly
changes dependent on the chosen filter. In order to check for focus stability the full width half
maximum (FWHM) of stellar point spread functions (PSF) were measured in CTK images during a focus
series. Thereby, the camera focus was always optimized in the I-band, and a FWHM of 1.5 pixel could
be achieved in the center of the CTK field of view. For the same focus the FWHM of the CTK PSF only
slightly increases in all other filters, namely 1.62\,pixel in B, 1.63\,pixel in V, 1.64\,pixel in
R, and 1.66\,pixel in z$'$. Hence, the FWHM of the CTK-PSF only varies maximal by 11\,\% for the
same focus in all filters, and remains always below 2\,CTK pixel. Hence, a refocusing of the camera
after a filter change is not needed.

\begin{figure}[h!]
\resizebox{\hsize}{!}{\includegraphics[]{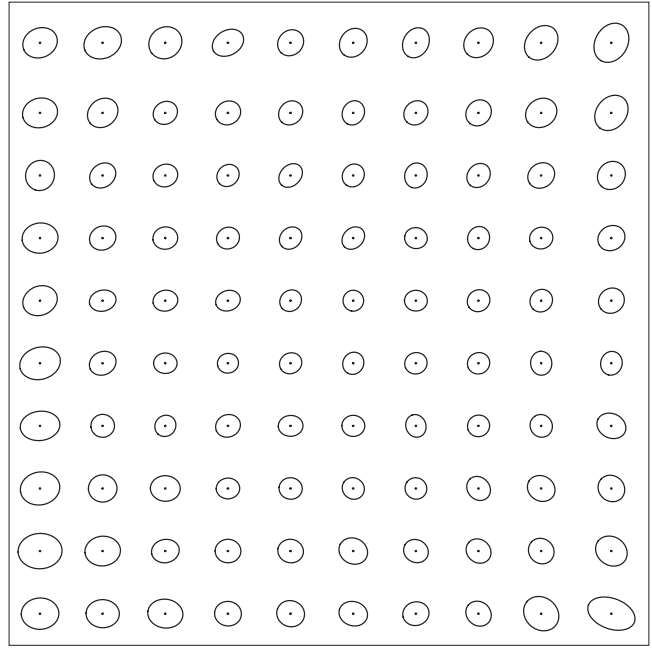}} \caption{This image shows the FWHM and
the shape of the CTK-PSF, dependent on position on the CTK detector, as measured through the V-band
filter. The smallest FWHM of the CTK-PSF is reached in the center of the detector and measures
1.5\,pixel. In this region the PSF is well radial symmetric. At the borders of the detector the
CTK-PSF appears more extended and also exhibit an elongated shape.} \label{imagequal1}
\end{figure}

\begin{figure}[h!]
\resizebox{\hsize}{!}{\includegraphics[]{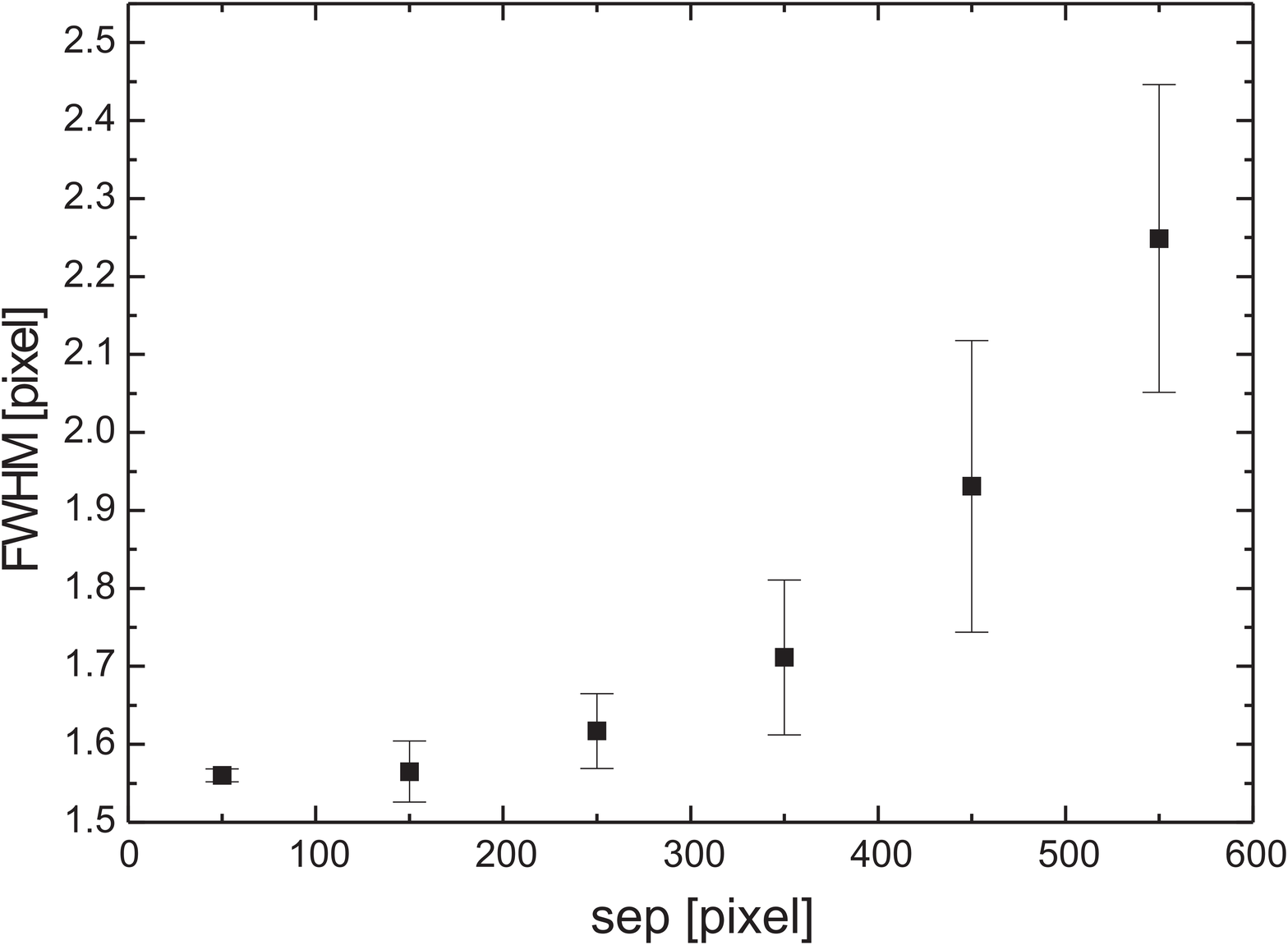}} \caption{The FWHM of the CTK-PSF for a
range of separation relative to the center of the detector. In order to determine the radial
variation of the PSF-FWHM, all detector cells, as shown in Fig.\,\ref{imagequal1}, are grouped
dependant on their separations to the center of the detector. The average, as well as the deviation
of the PSF-FWHM is derived for separation-bins of 100\,pixel. Within a separation of 300\,pixel
around the center the FWHM of the CTK-PSF does not show a significant variation, while it increases
for larger separations and exhibits larger deviations up to 9\,\% at the borders of the detector.}
\label{imagequal2}
\end{figure}

In order to determine the image quality of the CTK, images were taken which contain several hundred
of stars which are distributed over the whole field of view. For all detected point like sources in
the images, the FWHM as well as the elongation of the CTK-PSF is measured. The CTK field of view is
split then in cells which measure each 100$\times$100 pixel. The cells at the right border, as well
as at the top border of the detector include each 100$\times$124 pixel, while the cell in the
top-right corner of the detector measures 124$\times$124 pixel. The average of the FWHM and the
elongation of all point like sources, which are imaged in a cell, are derived. This yields the
average PSF-FWHM and elongation for each cell, and hence the image quality over the whole CTK field
of view.

The image quality of the CTK optics, as determined in the V-band after optimized focusing, is shown
in Fig.\,\ref{imagequal1}. In the center of the CTK field of view a PSF-FWHM of 1.5\,pixel is
achieved, and the PSF appears radial symmetric. The shape of the PSF becomes more elongated closer
to the borders of the detector. Furthermore, there is also an increase of the PSF-FWHM towards the
borders of the detector.

The averaged PSF-FWHM, dependant on the separation $sep$ to the center of the detector (pixel
[512,512]), is plotted in Fig.\,\ref{imagequal2}. Thereby, cells are grouped in bins of 100\,pixel
dependant an their separation to the center of the detector. Within a separation smaller than 100
pixel around the center the average of the PSF-FWHM is only 1.56$\pm$0.01\,pixel, i.e. a FWHM
deviation of only 0.6\,\%. The PSF-FWHM slightly increases up to 1.62$\pm$0.05\,pixel (deviation of
only 3\,\%) for a separation to the center between 200 and 300\,pixel. Hence, within 300\,pixel
around the center of the CTK detector the FWHM of the CTK-PSF is comparable, without a significant
variation. In contrast for wider separations the PSF-FWHM increases and also exhibits larger
deviations. Between 300 and 400\,pixel it already measures in average 1.7$\pm$0.1\,pixel,
1.9$\pm$0.2\,pixel for separations between 400 and 500\,pixel, and finally 2.3$\pm$0.2\,pixel
(deviation of 9\,\%) at the borders of the CTK detector ($sep>500$ pixel).

\section{CTK Quantum Efficiency}

\begin{figure}[h]
\resizebox{\hsize}{!}{\includegraphics[]{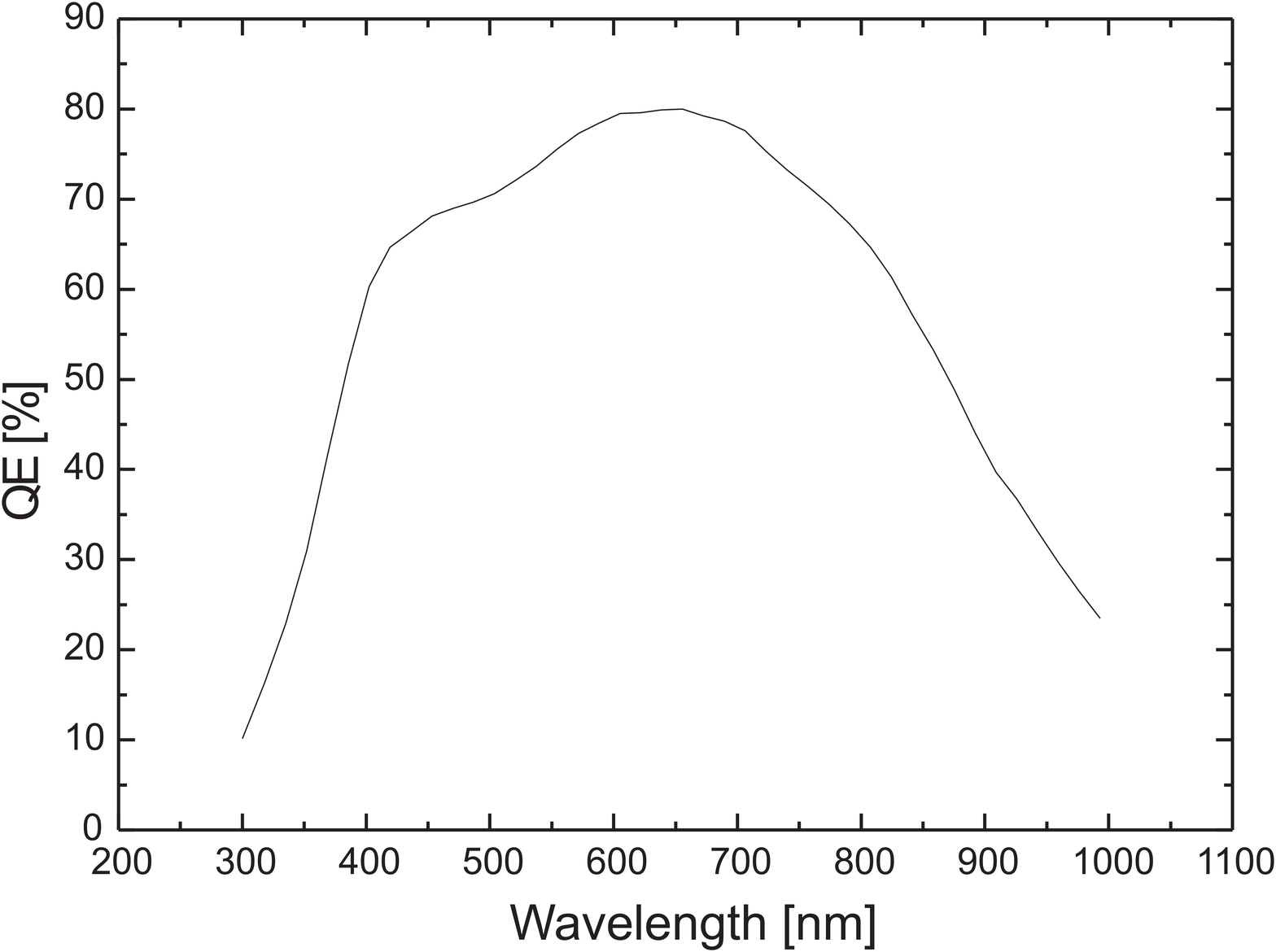}} \caption{The quantuum efficiency of the CTK
detector, measured at a temperature of 20$^{\circ}$C. Due to its back-illumination and the
antireflection coating a quantum efficiency up to 80\,\% is reached.} \label{qe}
\end{figure}

Because of its back-illumination as well as its antireflection coating the CTK detector exhibits a
high quantum efficiency (QE) which is illustrated in Fig.\,\ref{qe} as measured at a temperature of
20$^{\circ}$C. A maximum of 80\,\% is achieved at 650\,nm, and the QE remains above 70\,\% between
490\,nm and 770\,nm.

\section{CTK Detection Limits}

In order to determine the CTK detection limits standard stars were observed at an elevation of
about 45\,$^{\circ}$ (airmass 1.4) during dark time (no moon). The integration time for each
standard was always 1\,min with optimized focus for each filter (1.5\,pixel of PSF-FWHM). The
derived $S/N=3$ detection limits for all CTK filters are summarized in Tab.\,\ref{ctklimits}. The
highest sensitivity is reached in V- and R-band consistent with the quantum efficiency of the
detector, which reaches maximal values at the central wavelength of these filters. After 1\,min of
integration time stars with $V=R=17.2$\,mag can be detected at $S/N=3$.

\begin{table}[h!] \centering\caption{CTK detection limits ($S/N=3$) for 60\,s of integration time
during dark time (no moon). The limits are determined after an optimized focusing in each filter.}
\label{ctklimits}\begin{tabular}{ll|ll|ll}\hline
Filter & mag & Filter & mag & Filter & mag\\
\hline
B      & 16.4 & R      & 17.2 & z$'$   & 16.4\\
V      & 17.2 & I      & 16.8 &        &     \\
\hline\hline
\end{tabular}
\end{table}

\acknowledgements{I want to thank the staff of the mechanic and electronic division of the faculty
for physics and astronomy at the University Jena. Furthermore, I want to thank W. Pfau for all his
help and assistance during the preparation and installation phase of the CTK. This publication
makes use of data products from the Two Micron All Sky Survey, which is a joint project of the
University of Massachusetts and the Infrared Processing and Analysis Center/California Institute of
Technology, funded by the National Aeronautics and Space Administration and the National Science
Foundation, as well as the SIMBAD and VIZIER databases, operated at CDS, Strasbourg, France.}

\newpage

\begin{appendix}

\section{CTK Images}

\begin{figure}[h]
\resizebox{\hsize}{!}{\includegraphics[]{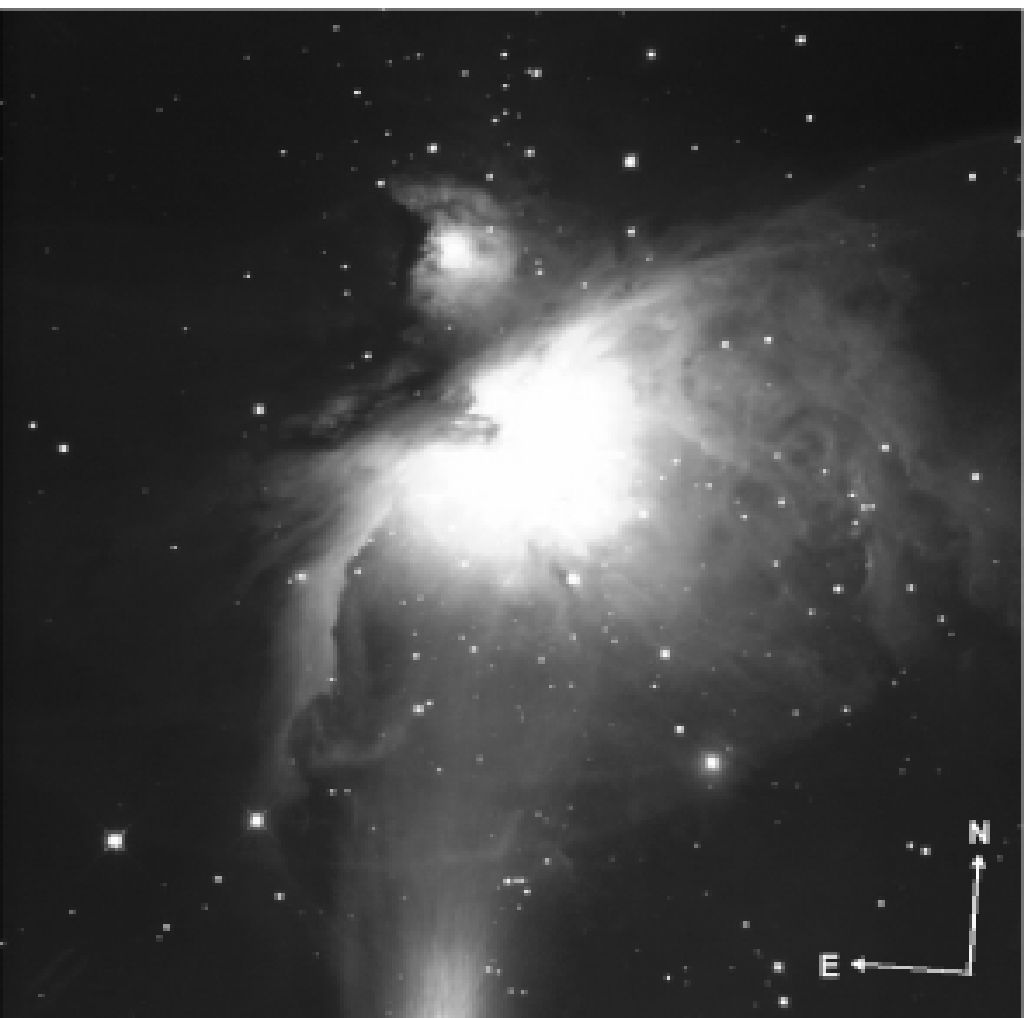}} \caption{The CTK \textsl{First Light}
taken on March 2th 2006. This image shows the Orion Nebula (M42) in the R-band. Seven CTK images
were taken with integration times between 30\,s and 300\,s, all averaged to the image shown here.
The achieved total integration time is 1050\,s. The image shows the full CTK field of view, which
measures 37.7$'$$\times$37.7$'$.} \label{firstlight}
\end{figure}

\begin{figure}[h!]
\resizebox{\hsize}{!}{\includegraphics[]{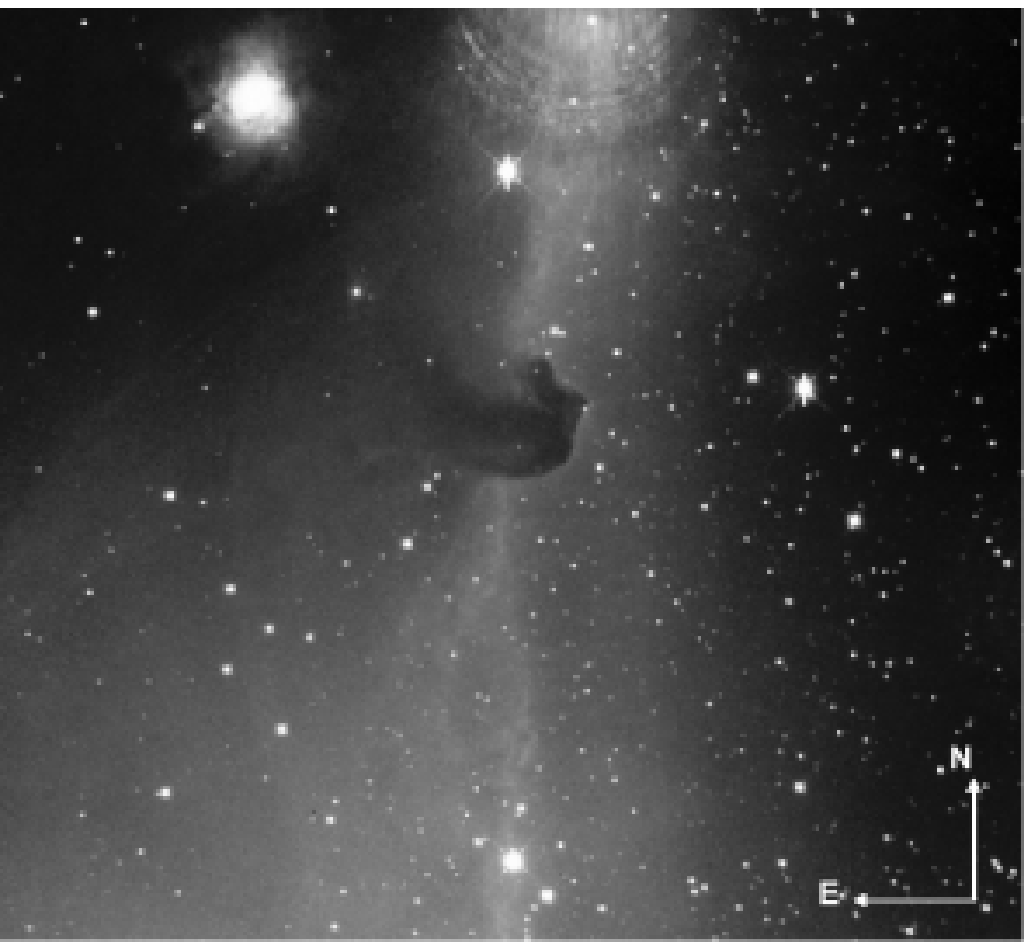}} \caption{The Horsehead Nebula (Barnard 33) in the
constellation Orion, imaged with the CTK in R-band. This image is a display of the whole CTK field
of view and covers 36.6$'$$\times$33.4$'$. Seven CTK images, each with an integration time of
120\,s, were averaged to the image shown here.} \label{hn}
\end{figure}

\begin{figure}[h]
\resizebox{\hsize}{!}{\includegraphics[]{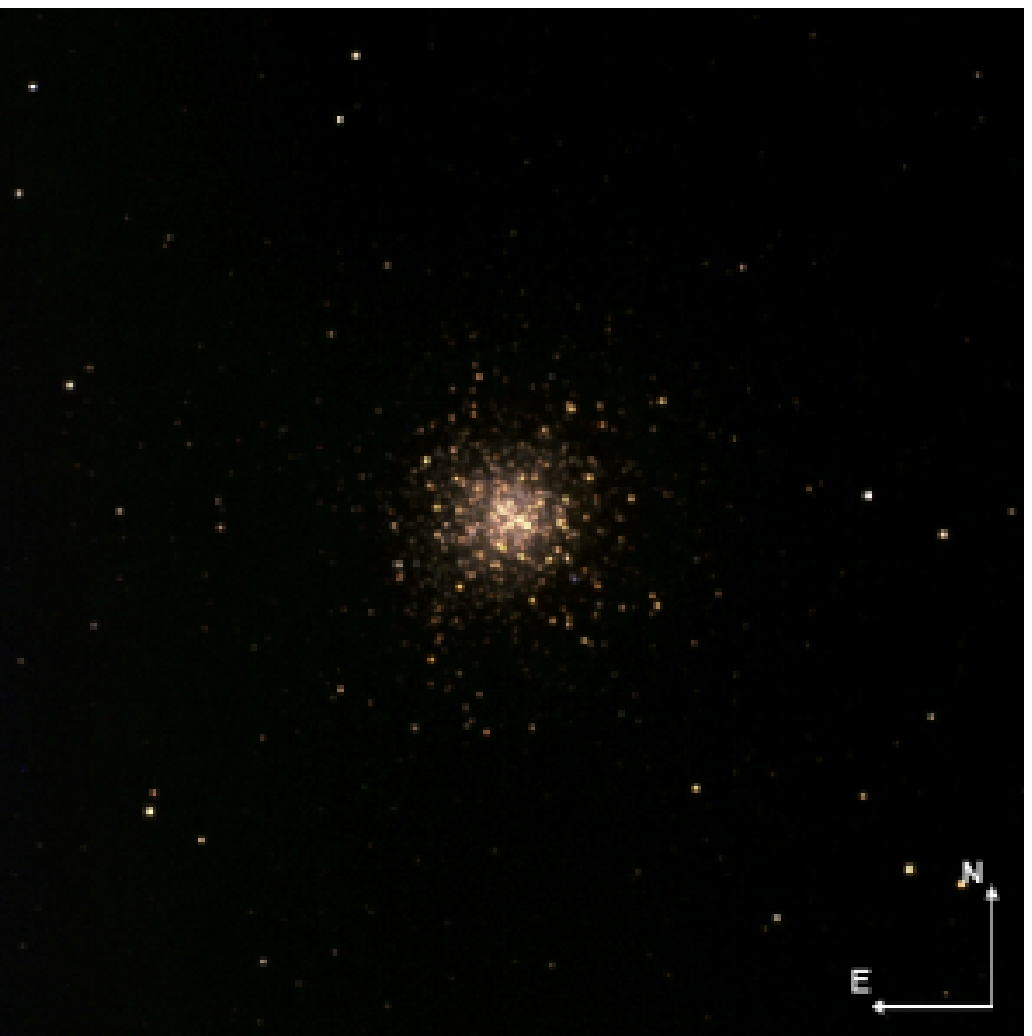}} \caption{This CTK image is a BVR-composite of
the globular cluster M13 in the constellation Hercules. The image shows the full CTK field of view,
which measures 37.7$'$$\times$37.7$'$. In the B-band three CTK images, each with an integration
time of 120\,s, were taken and averaged. Two 120\,s images were taken and averaged in the V- and
R-band, respectively.} \label{m13}
\end{figure}

\end{appendix}

\end{document}